\documentclass[11pt,twocolumn]{article}

\usepackage[T1]{fontenc}
\usepackage{mathptmx}
\usepackage{microtype}
\usepackage{amsmath}
\usepackage{amssymb}
\usepackage[margin=1in]{geometry}
\usepackage{booktabs}
\usepackage{caption}
\usepackage{enumitem}
\usepackage{float}
\usepackage{graphicx}
\usepackage{cite}
\usepackage{hyperref}
\usepackage{placeins}
\usepackage{supertabular}
\usepackage{tabularx}
\usepackage{titlesec}
\usepackage{xcolor}
\usepackage{tikz}
\usetikzlibrary{arrows.meta,positioning,calc}
\usepackage{algorithm}
\usepackage{algpseudocode}
\floatstyle{ruled}
\restylefloat{algorithm}

\graphicspath{{figures/}}
\hypersetup{
    colorlinks=true,
    linkcolor=blue!55!black,
    urlcolor=blue!55!black
}
\setlist[itemize]{leftmargin=1.45em,itemsep=0.35em,topsep=0.25em}
\titlespacing*{\section}{0pt}{0.8ex plus 0.25ex minus 0.15ex}{0.25ex}
\titlespacing*{\subsection}{0pt}{0.65ex plus 0.2ex minus 0.1ex}{0.2ex}

\title{Plate-Local River Generation: Deterministic, Terrain-Aware, Downhill-Flowing Rivers for Infinite Procedural Worlds}
\author{Michael K. Davis III, Independent Researcher, USA}

\makeatletter
\renewcommand{\@maketitle}{%
  \begin{center}
    \rule{\linewidth}{0.1pt}\\[1.2ex]
    {\LARGE\bfseries\@title\par}
    \vspace{0.5ex}
    \rule{\linewidth}{0.1pt}
  \end{center}
  \vspace{0.1ex}
  \begin{center}
    {\large\@author\par}
    \vspace{0.1ex}
    {\small\href{mailto:mikedavismk3@icloud.com}{\texttt{mikedavismk3@icloud.com}}\par}
  \end{center}
  \begin{center}
    \includegraphics[width=\textwidth]{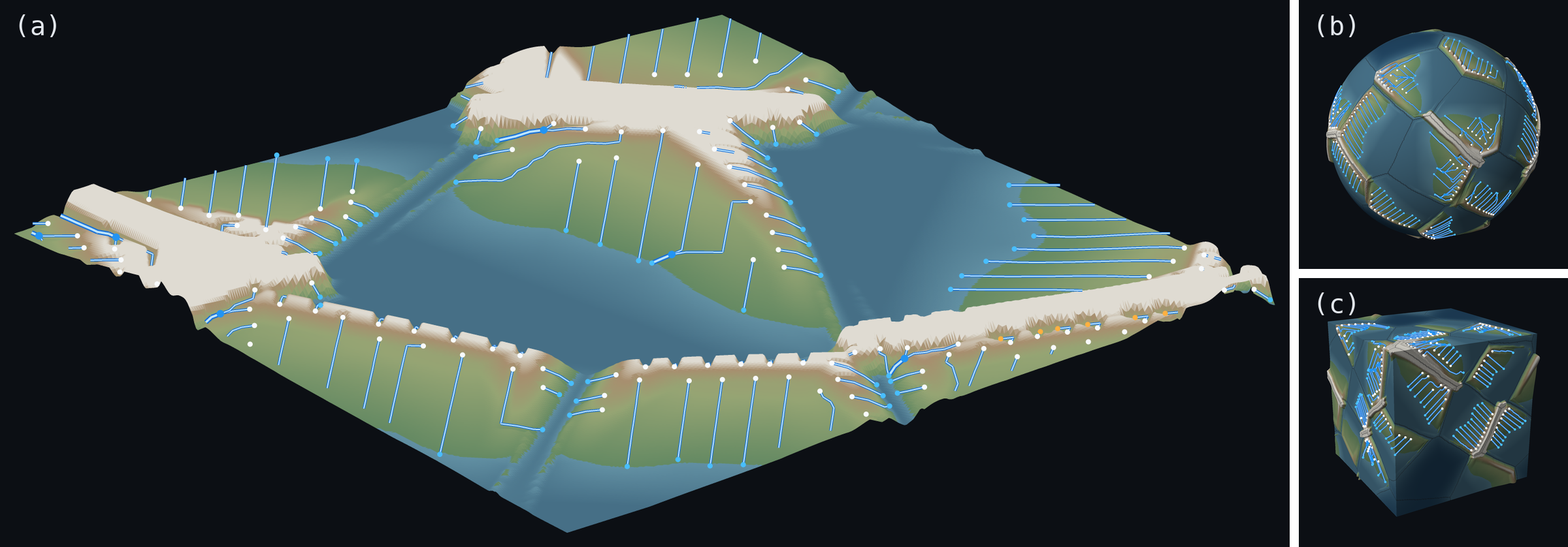}
    \captionof{figure}{Plate-local river networks generated on demand. (a) A regional view of the flat infinite-world implementation evaluated in Section~\ref{sec:reference-configuration}, in which ten plate domains independently generate terrain-aware, downhill-flowing river networks without global preprocessing. (b) Sphere and (c) cube planets from the finite-world toy implementations, each with 40 plates; rivers are cached for the plates within $55^\circ$ of the view direction, each built the first time its plate faces the camera. River geometry is depth-tested against the terrain, hiding channels occluded by ridges. White dots indicate source nodes, light blue dots sea outlets, darker blue dots confluences, and yellow dots lakes at local minima.}
    \label{fig:global-continent-rivers}
  \end{center}
  \vspace{1.0em}
}
\makeatother

\begin{document}

\maketitle

\begin{abstract}
Existing procedural terrain generation methods produce coherent, downhill-flowing river networks only on fixed domains, whereas approaches that extend to infinite worlds typically sacrifice network coherence or downhill-flow guarantees. This work presents a deterministic, terrain-aware, downhill-flowing river algorithm for infinite procedural worlds. Tectonic plate seeds define both continuous terrain fields and corresponding bounded Voronoi-like plate domains, so localized river preprocessing can operate directly on the geometric domain without sampling terrain fields for the purpose of locating plate boundaries. Within each plate, the terrain field is sampled on a low-resolution grid, and rivers are traced downhill from highland source nodes to coastlines, local minima, or existing confluences. River heights are then monotonically adjusted along each path so that every stored segment descends. The resulting river network is generated and cached lazily per plate, so each plate is processed independently of its neighbors. During full-resolution terrain sampling, the distance to the cached river network is computed exactly, and river height is interpolated along nearby segments and blended with the locally evaluated terrain on demand, without caching the blended result itself. Because the warp of the plate borders is bounded below the river border margin, this guarantees downhill flow along every river channel of the final terrain. Together, these design choices make terrain-aware river generation practical for interactive procedural worlds by supporting efficient on-demand evaluation, as demonstrated by the performance analysis presented in this work.
\end{abstract}

\section{Introduction}

Procedural, deterministic, locally generatable terrain is well suited to large worlds because the terrain can be generated in independent chunks that remain seamless with neighboring chunks. However, large-scale features such as river networks may depend on terrain relationships that extend far beyond the chunk currently being generated.

This work resolves that tension by using tectonic plates as an intermediate preprocessing domain---large enough to contain a plate's regional drainage structure, but small enough to generate and cache independently of the rest of the world. A plate's river network can be built directly from a geometric representation of the plate's domain, which itself does not require sampling the terrain fields to find its borders. Within this geometric domain, a low-resolution sample of the terrain fields is taken to guide downhill river generation. Because rivers are generated independently for each plate, they do not cross plate boundaries. This limitation is mitigated by constructing the continuous terrain fields so that oceans, plateaus, and mountain ranges naturally discourage cross-boundary drainage.

The key contributions of this work are:
\begin{itemize}
    \item dual plate representation for terrain sampling and river preprocessing, described in Sections~\ref{sec:dual-plate-representation}--\ref{sec:geometric-representation};
    \item lazy local river generation and caching, described in Section~\ref{sec:lazy-per-plate-river-cache}.
\end{itemize}

\section{Related Works}

Related works for fixed domains generally fall into two broad categories: systems that construct rivers first, then the terrain is constructed around them; and systems that construct the terrain first, then the rivers are constructed along this terrain. Learned terrain generation offers a novel approach to world generation in general, and can be extended to infinite domains~\cite{goslin2026}.

\subsection{River-First Generation} 

River-first approaches generate river network structures before the surrounding terrain~\cite{derzapf2011,genevaux2013,fischer2022}. By treating rivers as the primary organizing structure, these methods guarantee coherent, downhill-flowing rivers while allowing explicit control over river placement, connectivity, and slope. However, because terrain is constructed around a predefined river network, large-scale geological processes such as tectonic uplift and mountain building are secondary to river constraints. Furthermore, these approaches require preprocessing over a fixed domain and are not readily extendable to infinite procedural worlds.
\subsection{Terrain-First Generation} 

Terrain-first approaches first generate terrain before deriving river networks from the resulting topology. Batbold and Ziegelmeier~\cite{batbold2026} demonstrate that coherent river networks can be derived from procedurally generated terrain using graph-based flow routing. The present work builds on this terrain-first philosophy with river generation derived from a low-resolution sample of terrain fields, which is localized to plate domains. This enables river generation without preprocessing the entire domain.

\subsection{Learned Generation} 

InfiniteDiffusion~\cite{goslin2026} resolves the infinite-domain limitation of both river-first and terrain-first methods. Current limitations of diffusion-based terrain generators include limited explicit control over large-scale terrain features, such as continent placement, mountain range orientation, or coastline shape and no guarantee of coherent, downhill-flowing river networks. These limitations are not necessarily fundamental to diffusion models, but do remain unaddressed. 
\section{Plate-Local River Generation}

The following subsections define deterministic plate seeds and their dual pointwise and geometric representations, then describe plate-local terrain sampling, river tracing, caching, and on-demand composition. Together, these stages preserve terrain-aware downhill drainage without a global preprocessing pass. For applications to finite, closed-shaped worlds such as planet domains, we recommend slight modifications to this system, which we provide in this section alongside the infinite application.

\begin{figure*}[t]
    \centering
    \resizebox{\textwidth}{!}{%
    \begin{tikzpicture}[
        node distance=0.36cm,
        box/.style={
            draw,
            rounded corners=2pt,
            align=center,
            minimum height=0.88cm,
            text width=2.45cm,
            font=\small,
            fill=white
        },
        start/.style={box,fill=green!15},
        output/.style={box,fill=red!10},
        stage/.style={box,font=\small\bfseries},
        lane/.style={font=\small\bfseries,align=left},
        arrow/.style={-{Stealth[length=2.2mm]},thick},
        line/.style={thick}
    ]
        \node[start] (chunk) {Terrain view or\\chunk request};
        \node[stage,right=of chunk] (seeds) {\textbf{3.1}\\Deterministic\\Tectonic Plate\\Positions};

        \node[stage,anchor=west] (domain) at ($(seeds.east)+(0.9cm,1.25cm)$) {\textbf{3.4}\\Geometric\\Representation};
        \node[stage,right=of domain] (grid) {\textbf{3.5}\\Low-Resolution\\Plate Grid};
        \node[stage,right=of grid] (trace) {\textbf{3.6}\\River Node \&\\Segment Generation};
        \node[stage,right=of trace] (snapshot) {\textbf{3.7}\\Lazy Per-Plate\\River Cache};

        \node[stage,anchor=west] (platefields) at ($(seeds.east)+(0.9cm,-1.25cm)$) {\textbf{3.3}\\Pointwise\\Representation};

        \node[stage,text width=2.9cm,anchor=west] (composition) at ($(snapshot.east |- seeds.east)+(0.9cm,0)$) {\textbf{3.8}\\On-Demand Final\\Terrain Composition};
        \node[output,right=of composition] (voxel) {Locally generated\\voxel terrain};

        \node[lane,anchor=south west] at ([yshift=0.08cm]domain.north west) {Plate-local river cache build};
        \node[lane,anchor=north west] at ([yshift=-0.08cm]platefields.south west) {On-demand terrain evaluation};

        \draw[arrow] (chunk) -- (seeds);
        \coordinate (split) at ($(seeds.east)+(0.45cm,0)$);
        \draw[line] (seeds.east) -- (split);
        \draw[arrow] (split) |- (domain.west);
        \draw[arrow] (split) |- (platefields.west);

        \draw[arrow] (domain) -- (grid);
        \draw[arrow] (grid) -- (trace);
        \draw[arrow] (trace) -- (snapshot);

        \coordinate (merge) at ($(composition.west)+(-0.45cm,0)$);
        \draw[line] (snapshot.east) -| (merge);
        \draw[line] (platefields.east) -| (merge);
        \draw[arrow] (merge) -- (composition.west);
        \draw[arrow] (composition) -- (voxel);
    \end{tikzpicture}%
    }
    \caption{Pipeline overview, generalized for various world types---infinite or finite, and landscape structures---voxels or mesh.}
    \label{fig:pipeline-overview}
\end{figure*}
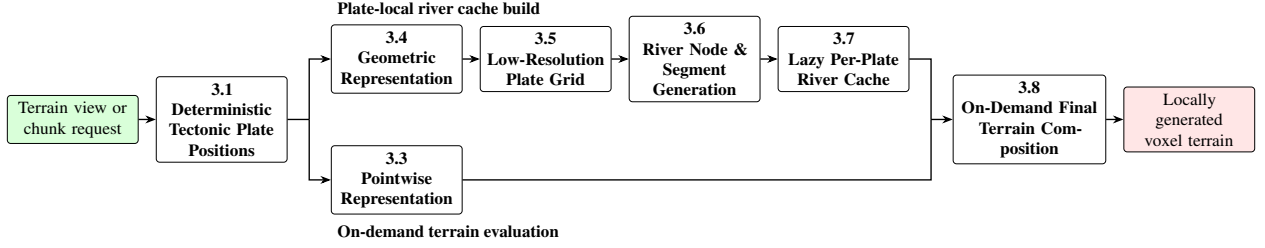

Figure~\ref{fig:pipeline-overview} summarizes this pipeline, separating the per-plate cache build from on-demand terrain evaluation. Algorithms 1 and 2 detail the complete procedures for building plate-local river caches and evaluating terrain on demand.

\begingroup
\footnotesize
\noindent\rule{\linewidth}{1pt}\par\smallskip
\refstepcounter{algorithm}
\noindent\textbf{Algorithm~\thealgorithm.} \textit{Plate-Local River Cache Build}\label{alg:cache-build}\par\smallskip
\noindent\rule{\linewidth}{0.6pt}\par\smallskip
\begin{algorithmic}[1]
\Require Plate $i$, seed positions $\{\mathbf{c}_j\}$, surface $\mathcal{S}$
\Ensure River network $\mathcal{R}_i$
\Statex \textit{Geometric plate domain (Section 3.4)}
\State Initialize bounding volume $B_i$ around $\mathbf{c}_i$
\For{each neighboring plate $j \in \mathcal{N}_i$}
    \State $\mathbf{m}_{ij} \gets \frac{1}{2}(\mathbf{c}_i+\mathbf{c}_j)$;
    $\mathbf{n}_{ij} \gets \mathbf{c}_j-\mathbf{c}_i$
    \State Clip $B_i$ by $(\mathbf{p}-\mathbf{m}_{ij})\cdot\mathbf{n}_{ij}\leq 0$
\EndFor
\State $\Omega_i^{\mathcal{S}} \gets B_i \cap \mathcal{S}$
\Statex \textit{Low-resolution plate grid (Section 3.5)}
\For{each $(u,v)$ with $\mathbf{x}_{uv}\in\Omega_i^{\mathcal{S}}$}
    \State $h_{uv}\gets H_{\mathrm{plan}}(\mathbf{x}_{uv})$
    \State Mark $(u,v)$ border-safe and/or lake-safe from $\operatorname{dist}(\mathbf{x}_{uv},\partial\Omega_i^{\mathcal{S}})$
\EndFor
\Statex \textit{River node and segment generation (Section 3.6)}
\State $\mathcal{R}_i\gets\emptyset$
\For{each border-safe source candidate, in descending height order}
    \If{$K$ paths are accepted}
        \State \textbf{break}
    \EndIf
    \If{the candidate is within $s_{\min}$ of an existing river}
        \State \textbf{continue}
    \EndIf
    \Repeat
        \If{at sea level}
            \State Mark a sea outlet; \textbf{break}
        \EndIf
        \If{a lower existing node is within reach}
            \State Join it as a confluence; \textbf{break}
        \EndIf
        \State Step to the lowest border-safe, non-crossing lower sample
    \Until{no lower sample exists or $n_{\max}$ steps are taken}
    \State At a dead end: mark a lake if lake-safe, else discard the path
    \State Add the accepted path's segments to $\mathcal{R}_i$
\EndFor
\State Monotonically adjust river heights along each path
\State Assign Strahler orders; cache grid, nodes, segments, and paths
\end{algorithmic}
\smallskip\noindent\rule{\linewidth}{1pt}
\endgroup

\begingroup
\footnotesize
\medskip\noindent\rule{\linewidth}{1pt}\par\smallskip
\refstepcounter{algorithm}
\noindent\textbf{Algorithm~\thealgorithm.} \textit{On-Demand Terrain Sample}\label{alg:terrain-sample}\par\smallskip
\noindent\rule{\linewidth}{0.6pt}\par\smallskip
\begin{algorithmic}[1]
\Require Position $\mathbf{x}$, seed positions $\{\mathbf{c}_j\}$, river cache
\Ensure Final height $H_{\mathrm{final}}(\mathbf{x})$
\Statex \textit{Plate assignment (Section 3.3)}
\State $\tilde{\mathbf{x}}\gets s_p\mathbf{x}+\mathbf{w}(\mathbf{x})$
\State $i\gets\operatorname*{arg\,min}_{j\in\mathcal{A}}
\|\tilde{\mathbf{x}}-\mathbf{c}_j\|^2$ \Comment{Owning plate; local $3\times3$ for 2D or $3\times3\times3$ for 3D lattice search if $\mathcal{A}=\mathbb{Z}^d$}
\Statex \textit{Cache lookup and lazy build (Section 3.7)}
\If{the cache for plate $i$ is absent}
    \State Execute Algorithm~\ref{alg:cache-build} for plate $i$
\EndIf
\Statex \textit{Pointwise terrain fields (Section 3.3)}
\State Evaluate $C$, $W_{\mathrm{land}}$, $W_{\mathrm{mountain}}$,
$W_{\mathrm{plateau}}$, and $W_{\mathrm{island}}$ at $\mathbf{x}$
\State Compute $H_{\mathrm{pre}}(\mathbf{x})$ from the weighted fields
\Statex \textit{River query (Section 3.7)}
\If{$\mathcal{R}_i=\emptyset$}
    \State \Return $H_{\mathrm{pre}}(\mathbf{x})$
\EndIf
\State $\delta_{\mathrm{river}}(\mathbf{x})\gets
\min_{s\in\mathcal{R}_i}\delta_s(\mathbf{x})$
\Comment{Scan all segments of plate $i$; lake disks included}
\State Blend $h_s(\mathbf{x})$ over nearby $s$ into $H_{\mathrm{river}}(\mathbf{x})$
\Statex \textit{Terrain composition (Section 3.8)}
\State $\lambda_{\mathrm{river}}(\mathbf{x})\gets
\left[\min\!\left(1,\frac{[\delta_{\mathrm{river}}(\mathbf{x})]_+}
{\omega_{\mathrm{river}}}\right)\right]^{\alpha_{\mathrm{river}}}$
\State $H_{\mathrm{final}}(\mathbf{x})\gets
\lambda_{\mathrm{river}}(\mathbf{x})H_{\mathrm{pre}}(\mathbf{x})$
\Statex \hspace{\algorithmicindent}$\quad+
(1-\lambda_{\mathrm{river}}(\mathbf{x}))H_{\mathrm{river}}(\mathbf{x})$
\State \Return $H_{\mathrm{final}}(\mathbf{x})$
\end{algorithmic}
\smallskip\noindent\rule{\linewidth}{1pt}
\endgroup

\subsection{Deterministic Tectonic Plate Positions}

Each tectonic plate is defined by a deterministic seed position. The same construction can be used for finite and infinite worlds: finite worlds use a bounded set of plate identifiers, while infinite worlds generate plate identifiers from lattice cells on demand. In both cases, terrain field samples are assigned to the nearest deterministic seed position, producing Voronoi-like plate regions~\cite{aurenhammer1991,okabe2000}.

For an infinite world, plate seeds may be generated as one deterministic point per integer lattice cell, following the same nearest-feature-point structure used by cellular texture bases~\cite{worley1996},
\begin{equation}
    \mathbf{c}_i =
    \mathbf{i}+\tfrac{1}{2}\mathbf{1}+r\,\boldsymbol{\eta}(\mathbf{i}),
    \qquad \mathbf{i}\in\mathbb{Z}^d,
\end{equation}
where $i$ is the plate identifier associated with integer lattice-cell coordinate $\mathbf{i}$, $\boldsymbol{\eta}(\mathbf{i})\in[-1,1]^d$ is a deterministic per-axis hash offset, and $r$ bounds the jitter in cell widths. Seeds are expressed in \emph{plate space}, in which world coordinates are multiplied by the plate scale $s_p$ (Equation~\eqref{eq:warped-plate-space}), so one lattice cell spans $1/s_p$ world units. For this present implementation, $\boldsymbol{\eta}(\mathbf{i})$ remains an offset on the 2D plane. Considering an additional offset in 3D space will result in continental landmasses that are less uniform in size, which may be preferred. 

For a finite world, with a closed geometry and $N$ plates, a golden-angle sequence gives approximately even seed directions~\cite{saff1997,swinbank2006}:
{\small
\begin{gather}
    \theta_i = i\pi(3-\sqrt{5}), \qquad
    y_i = 1-\frac{2i+1}{N}, \qquad
    r_i = \sqrt{1-y_i^2}, \\
    \mathbf{q}_i =
    \left(r_i\cos\theta_i,\; y_i,\; r_i\sin\theta_i\right),
    \\
    \mathbf{c}_i = s_p\Pi_{\mathcal{S}}(\mathbf{q}_i)+\boldsymbol{\eta}_i,
    \quad 0 \le i < N,
\end{gather}
}
Here $\Pi_{\mathcal{S}}$ projects the unit direction onto the chosen finite world surface, and $\boldsymbol{\eta}_i$ is a deterministic jitter applied to the finite plate center.

\subsection{Dual Tectonic Plate Representation}
\label{sec:dual-plate-representation}

The tectonic seed positions are used in two forms: a \emph{pointwise representation}, defined by continuous coordinate-based fields for terrain evaluation, and a \emph{geometric representation}, defined by a bounded non-rasterized domain for regional river preprocessing. The same seed position is used to determine both representations for that given plate.

\begin{figure}[t]
    \centering
    \includegraphics[width=0.9\linewidth]{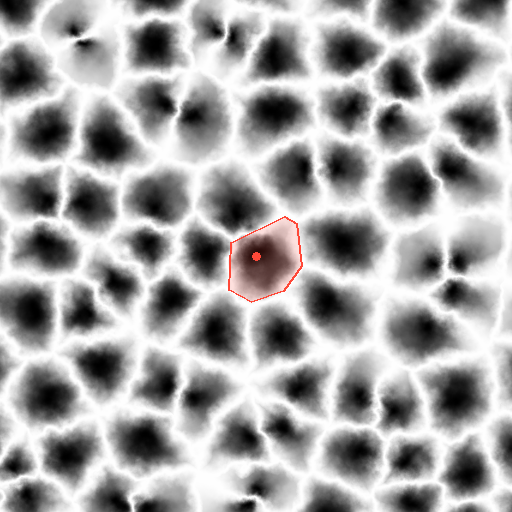}
    \caption{Dual plate representations. A continuous terrain field (grayscale) and a corresponding geometric plate domain (red) are independently derived from the same tectonic plate seed.}
    \label{fig:plate-geometry}
\end{figure}

Figure~\ref{fig:plate-geometry} shows both representations derived from the same plate seed.

\subsection{Pointwise Representation}

During chunk generation, terrain is evaluated directly from world-space coordinates by comparing plate seed positions and deriving continuous scalar and vector weight fields from plate ownership, plate type, border distance, and drift direction. These weight fields control land, mountain, plateau, and island chain placement. Noise warping and surface roughness are applied to produce more organic continent shapes~\cite{perlin1985}.

The primary purposes of the pointwise representation are to generate terrain features that appear natural, and to generate terrain features with natural barriers, such as mountains, plateaus, and oceans, at plate borders, providing a natural motive for localizing rivers to each plate and prohibiting plate boundary crossings. Otherwise, the pointwise representation is not essential to this system, and the continuous terrain fields can be constructed to disregard the plate boundaries entirely.

\begin{figure*}[t]
    \centering
    \includegraphics[width=\textwidth]{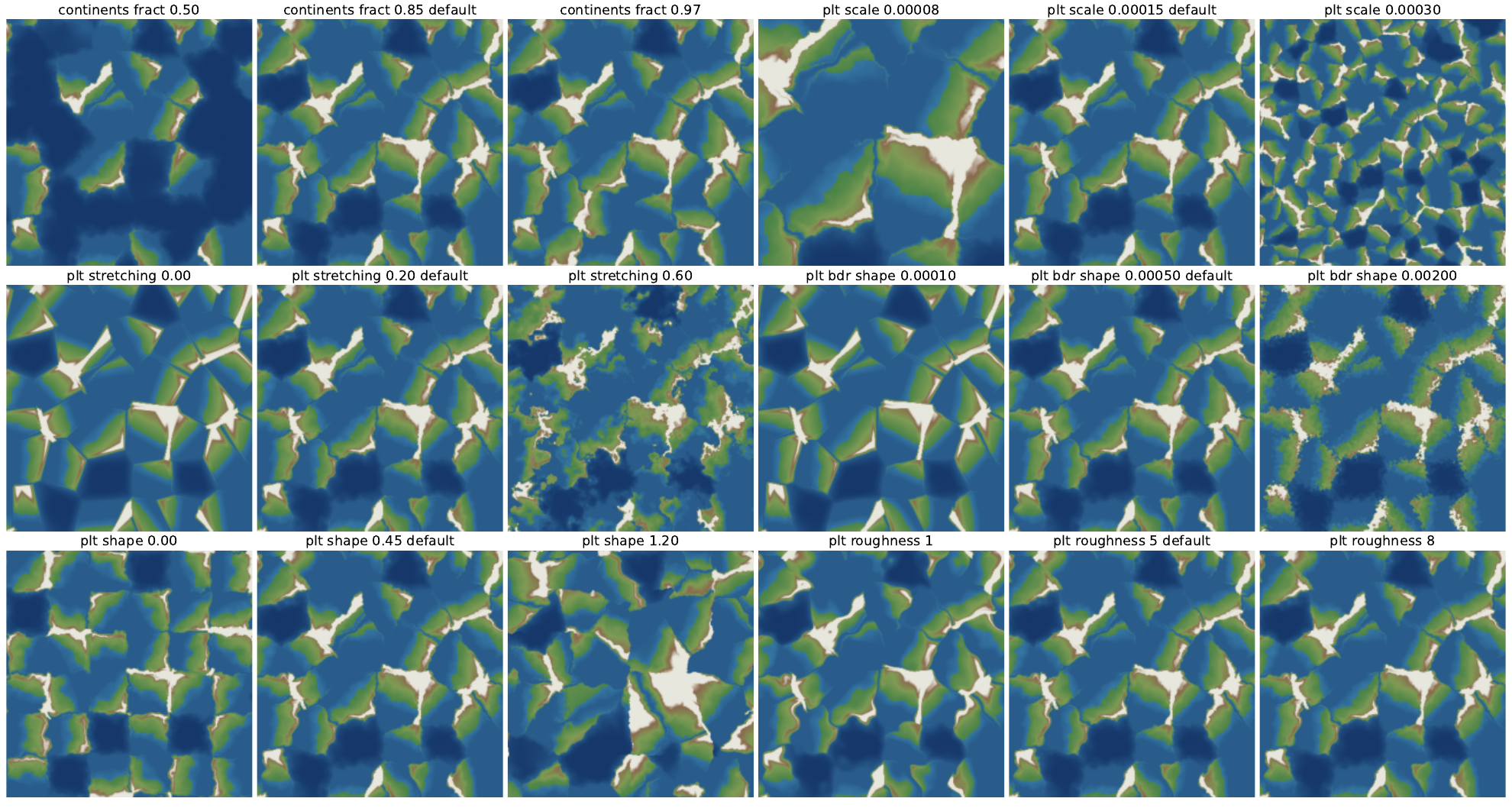}
    \caption{Illustrative top-down height-field renders from an earlier prototype of the pointwise terrain, showing how individual plate-tessellation parameters change the terrain. Each panel varies one parameter (low, prototype default, high) for one fixed seed, with all other parameters at the prototype's defaults; panel labels give the prototype's parameter names and values. The prototype's terrain and defaults differ from the implementation evaluated in the results section, so these renders are not comparable to Figure~\ref{fig:global-continent-rivers}. Row 1: oceanic/continental threshold, then plate scale $s_p$. Row 2: warp amplitude, then warp spatial frequency. Row 3: plate-seed jitter (zero jitter collapses to a regular lattice), then warp octave count.}
    \label{fig:birds-eye-grid}
\end{figure*}

As a qualitative illustration, Figure~\ref{fig:birds-eye-grid} shows how several plate-tessellation parameters shape the terrain, using renders from an earlier prototype of the pointwise representation. 

\paragraph{Pointwise Plate.} For a world-space point $\mathbf{x}$, the pointwise plate fields are evaluated in warped plate space,
\begin{equation}
    \label{eq:warped-plate-space}
    \tilde{\mathbf{x}} = s_p \mathbf{x} + \mathbf{w}(\mathbf{x}),
\end{equation}
where $\mathbf{w}(\mathbf{x}) = a_w\big(n_1(\mathbf{x}),\,n_2(\mathbf{x})\big)$ is a deterministic noise warp built from two normalized fractal noise fields $n_1,n_2\in[-1,1]$ with amplitude $a_w$ in plate cells, so $\|\mathbf{w}\|\le\sqrt{2}\,a_w$. Note that $\mathbf{w}$ cannot be represented in the geometric representation of the same plate. Given deterministic plate centers $\mathbf{c}_i$, the owning plate is the nearest center in this space:

\begin{equation}
    i = \operatorname*{arg\,min}_{j\in\mathcal{A}}
    \left\|\tilde{\mathbf{x}} - \mathbf{c}_j\right\|^2 .
\end{equation}
Here $\mathcal{A}$ is the set of possible plate identifiers. For a finite world, $\mathcal{A} = \{1,\ldots,N\}$ and each identifier names one generated plate seed. For an infinite world, $\mathcal{A} = \mathbb{Z}^d$ and each identifier is an integer lattice-cell coordinate that generates one seed on demand. Rather than searching all of $\mathbb{Z}^d$, the implementation restricts this arg min to the $3\times3$ for 2D or $3\times3\times3$ for 3D neighborhood of lattice cells surrounding $\tilde{\mathbf{x}}$'s own cell, following the same bounded-jitter neighborhood search used for cellular texture bases~\cite{worley1996}. This restriction is exact, independent of the warp $\mathbf{w}$, provided the per-axis hash offset is bounded to at most $r$ cell-widths from the cell center: for a query point at a corner shared by four cells, each of those four seeds can be pushed no farther than $(0.5+r)\sqrt{2}$ from the corner, while a seed two cells away along either axis can be pulled as close as $\sqrt{(1.5-r)^2+(r-0.5)^2}$. These two distances are equal at $r=1/3$, so bounding the hash offset to $r\le1/3$ guarantees the neighborhood always contains the true nearest seed; the implementation enforces $r=1/3$.
Plate fields are controlled by the distance margin between the owning plate and a neighboring plate:
\begin{equation}
    d_j(\mathbf{x}) =
    \left\|\tilde{\mathbf{x}} - \mathbf{c}_j\right\|
    -
    \left\|\tilde{\mathbf{x}} - \mathbf{c}_i\right\|,
    \qquad j \ne i .
\end{equation}
Small values of $d_j$ indicate proximity to the boundary between the owning plate $i$ and neighbor plate $j$. When a terrain feature depends on the nearest eligible neighboring plate, the margin is minimized over a selected neighbor set $\mathcal{N}$:
\begin{equation}
    d_{\mathcal{N}}(\mathbf{x}) =
    \min_{j \in \mathcal{N}} d_j(\mathbf{x}) .
\end{equation}

\paragraph{Plate Drift.} To mimic the effects of tectonic plate drift, where land, mountains, and plateaus are largest near colliding plate borders, we define plate drift vector and scalar fields to guide the magnitudes of subsequent weight fields.

Each plate identifier is assigned a deterministic drift axis $\mathbf{a}_i$ by projecting a random vector $\mathbf{r}_i$ into the tangent plane of the terrain domain. For a non-flat world, such as a finite planet with a closed geometry, the plate center direction vector $\mathbf{c}_i$ from the world origin provides a stable reference normal:
\begin{align}
    \mathbf{a}_i &= \frac{\mathbf{c}_i \times \mathbf{r}_i}
    {\left\|\mathbf{c}_i \times \mathbf{r}_i\right\|}.
\end{align}
For a flat world, such as an infinite one, the constant reference normal $\mathbf{n}_0$ can be used instead:
\begin{align}
    \mathbf{a}_i &= \frac{\mathbf{n}_0 \times \mathbf{r}_i}
    {\left\|\mathbf{n}_0 \times \mathbf{r}_i\right\|}.
\end{align}
The drift direction at the terrain sample is then
\begin{align}
    \boldsymbol{\ell}(\mathbf{x}) &=
    \frac{\mathbf{a}_i \times \mathbf{n}(\mathbf{x})}
    {\left\|\mathbf{a}_i \times \mathbf{n}(\mathbf{x})\right\|},
\end{align}
where $\mathbf{n}(\mathbf{x})$ is the local surface normal, and $\boldsymbol{\ell}(\mathbf{x})$ is the local tangent drift direction used at the terrain sample. For a cube-shaped planet, the implementation uses the radial direction $\mathbf{x}/\|\mathbf{x}\|$ in place of the true face normal, which keeps $\boldsymbol{\ell}$ continuous across cube edges. Note that for a flat world, this greatly simplifies to a constant per-plate direction $\boldsymbol{\ell}(\mathbf{x}) = \mathbf{r}_i/\|\mathbf{r}_i\|$ with $\mathbf{r}_i \in \mathbb{R}^2$.

The scalar drift field, $\rho(\mathbf{x})$, measures signed progress along the local drift direction:
\begin{equation}
    \rho(\mathbf{x}) =
    \left(\tilde{\mathbf{x}} - \mathbf{c}_i\right)
    \cdot
    \boldsymbol{\ell}(\mathbf{x}) .
\end{equation}

\paragraph{Weights.} Land, island, and mountain feature weights use the same normalized beta-shaped border profile. For feature $k$ with amplitude $A_k$, width $\omega_k$, interior exponent $p_k$, border exponent $q_k$, and plate-coverage parameter $\gamma_k$,
\begin{gather}
    C_k =
    \left[
    \left(\frac{p_k}{p_k+q_k}\right)^{p_k}
    \left(\frac{q_k}{p_k+q_k}\right)^{q_k}
    \right]^{-1}, \\
    D_k(\mathbf{x}) =
    \max\left(0,\frac{\rho(\mathbf{x}) + 2\gamma_k - 1}{2\gamma_k}\right), \\
    z_k(\mathbf{x}) =
    \frac{d_{\mathcal{N}_k}(\mathbf{x})}{\omega_k}, \\
    W_k(\mathbf{x}) =
    \begin{cases}
    \begin{aligned}
    &A_k C_k D_k(\mathbf{x})z_k(\mathbf{x})^{p_k} \\
    &\quad{}\cdot\left(1-z_k(\mathbf{x})\right)^{q_k},
    \end{aligned}
    & 0 \le z_k(\mathbf{x}) \le 1, \\
    0, & \text{otherwise.}
    \end{cases}
\end{gather}
The coefficient $C_k$ normalizes the profile before amplitude scaling, and $D_k$ gates the feature by the drift field.

Plateaus are treated separately because, in this work, we define them as highlands arising from converging continental plates rather than a single boundary-distance profile. For a nearby continental pair $(i,j)$, let $\widehat{\mathbf{u}}_{ij}(\mathbf{x})$ denote the unit separation direction from plate $i$ to plate $j$ after projection into the local tangent plane. The closing response is measured by projecting each plate's tangent drift onto this direction:
{\small
\begin{gather}
    \kappa_{i\to j}(\mathbf{x}) =
    \begin{cases}
    \left(\mathbf{a}_i \times \mathbf{n}(\mathbf{x})\right)
    \cdot \widehat{\mathbf{u}}_{ij}(\mathbf{x}),
    & \text{if positive,} \\
    0, & \text{if zero or negative,}
    \end{cases} \\
    \kappa_{j\to i}(\mathbf{x}) =
    \begin{cases}
    -\left(\mathbf{a}_j \times \mathbf{n}(\mathbf{x})\right)
    \cdot \widehat{\mathbf{u}}_{ij}(\mathbf{x}),
    & \text{if positive,} \\
    0, & \text{if zero or negative.}
    \end{cases}
\end{gather}
}
These detect colliding plates. The weight itself is constructed as,
\begin{align}
    R_{\mathrm{plateau}}(\mathbf{x})
    ={}& O(\mathbf{x})
    \max_{(i,j)\in\mathcal{C}}
    B_{ij}(\mathbf{x})N_i(\mathbf{x}) \notag\\
    &{}\cdot \kappa_{i\to j}(\mathbf{x})
    \kappa_{j\to i}(\mathbf{x}) \notag\\
    &{}\cdot S\!\left(\frac{\kappa_{i\to j}(\mathbf{x})}{\eta}\right)
    S\!\left(\frac{\kappa_{j\to i}(\mathbf{x})}{\eta}\right), \\
    W_{\mathrm{plateau}}(\mathbf{x})
    ={}& A_{\mathrm{plateau}}
    S\!\left(g_{\mathrm{plateau}}R_{\mathrm{plateau}}(\mathbf{x})\right) \notag\\
    &{}\cdot \left(A_{\mathrm{land}}-W_{\mathrm{land}}(\mathbf{x})\right) \notag\\
    &{}\cdot \left(A_{\mathrm{mountain}}-W_{\mathrm{mountain}}(\mathbf{x})\right).
\end{align}
Here $\mathcal{C}$ is the set of continental plate pairs among the plates considered at $\mathbf{x}$ (the $3\times3$ lattice neighborhood for an infinite world), $S$ is a smoothstep function, and $\eta$ is the closing fade width. With bridge width $\omega_{b}$, $B_{ij}(\mathbf{x}) = 1-S\big((\|\tilde{\mathbf{x}}-\mathbf{c}_j\|-\|\tilde{\mathbf{x}}-\mathbf{c}_i\|)/\omega_b\big)$ fades with the border margin between plates $i$ and $j$ at $\mathbf{x}$, $N_i(\mathbf{x})$ applies the same fade to the margin between plate $i$ and the owning plate, and $O(\mathbf{x}) = S\big(d_{\mathrm{ocean}}(\mathbf{x})/\omega_b\big)$ suppresses plateau formation near oceanic boundaries, where $d_{\mathrm{ocean}}$ is the margin to the nearest oceanic plate. The gain $g_{\mathrm{plateau}}$ controls the plateau response ($g_{\mathrm{plateau}}=4$ in the implementation). The final plateau weight $W_{\mathrm{plateau}}$ is an amplitude-scaled smoothstep response to the closing response $R_{\mathrm{plateau}}$, suppressed where the land and mountain fields already dominate; the implementation clamps both suppression factors at zero.

Island chains are further confined to the leading edge of each plate and removed beneath plateaus:
\begin{equation}
    \begin{aligned}
    W_{\mathrm{island}} \leftarrow{}& W_{\mathrm{island}}
    \max\!\left(0,\,1+h_{\mathrm{sea}}-\frac{\rho(\mathbf{x})-\rho_0}{L}\right) \\
    &\cdot\left(1-\min(1,W_{\mathrm{plateau}})\right),
    \end{aligned}
\end{equation}
where $h_{\mathrm{sea}}$ is the sea level, $\rho_0 = 1-2\gamma_{\mathrm{land}}+h_{\mathrm{sea}}$, and $L = 2(\gamma_{\mathrm{island}}-\gamma_{\mathrm{land}})$. Oceanic plates receive none of the land, island, mountain, or plateau weights.

\paragraph{Pre-River Terrain Height Composition.} The pre-river terrain height $H_{\mathrm{pre}}(\mathbf{x})$ is the sum of a crust field and the land, island, mountain, and plateau weight fields:
\begin{equation}
    \label{eq:pre-river-height}
    \begin{aligned}
    H_{\mathrm{pre}}(\mathbf{x}) ={}& C(\mathbf{x}) + W_{\mathrm{land}}(\mathbf{x}) + W_{\mathrm{mountain}}(\mathbf{x}) \\
    &+ W_{\mathrm{plateau}}(\mathbf{x}) + n_{\mathrm{island}}(\mathbf{x})\,W_{\mathrm{island}}(\mathbf{x}),
    \end{aligned}
\end{equation}
where $n_{\mathrm{island}}(\mathbf{x})\in[0,1]$ is low-frequency noise that breaks island chains into individual islands. The crust field sets the height floor that distinguishes continental from oceanic plates. It is the constant $h_{\mathrm{cont}}$ on continental plates, while on oceanic plates it rises toward $h_{\mathrm{cont}}$ near continental borders, forming a continental shelf:
\begin{equation}
    C(\mathbf{x}) = h_{\mathrm{ocean}} + \left(h_{\mathrm{cont}}-h_{\mathrm{ocean}}\right)
    \left[1-\frac{d_{\mathrm{cont}}(\mathbf{x})}{\omega_{\mathrm{shelf}}}\right]_+^{p_{\mathrm{shelf}}},
\end{equation}
where $d_{\mathrm{cont}}$ is the margin to the nearest continental plate, $\omega_{\mathrm{shelf}}$ is the shelf width, $p_{\mathrm{shelf}}$ is the shelf exponent, and $[\cdot]_+=\max(0,\cdot)$.

\subsection{Geometric Representation}
\label{sec:geometric-representation}

The entire plate domain can be constructed geometrically from nearby seed positions without terrain field sampling. Because the noise warping applied to the weight fields cannot be reproduced geometrically, this domain is a straight-bordered approximation of the true plate footprint---close enough for river generation purposes.

The geometric plate domain is initialized as a bounded volume $B_i$ around the plate center and then clipped by the bisector planes between $\mathbf{c}_i$ and $\mathbf{c}_j$:
\begin{equation}
    \Omega_i =
    B_i \cap
    \bigcap_{j \ne i}
    \left\{
    \mathbf{p} :
    \left(\mathbf{p} - \mathbf{m}_{ij}\right)
    \cdot
    \mathbf{n}_{ij}
    \le 0
    \right\},
\end{equation}
where
\begin{align}
    \mathbf{m}_{ij} &= \frac{1}{2}\left(\mathbf{c}_i+\mathbf{c}_j\right), \\
    \mathbf{n}_{ij} &= \mathbf{c}_j-\mathbf{c}_i .
\end{align}
In practice the intersection runs over a finite neighbor set $j \in \mathcal{N}_i$. For an infinite world, the implementation uses the eight lattice cells surrounding plate $i$, and $B_i$ is the axis-aligned rectangle enclosing these nine seeds, padded by two cell widths. For a finite world, $\mathcal{N}_i$ contains every other plate. For a world surface $\mathcal{S}$, the plate-local preprocessing domain is the surface cross section of this clipped volume,
\begin{equation}
    \Omega_i^\mathcal{S} =
    \Omega_i \cap \mathcal{S}.
\end{equation}
For a flat world, $\Omega_i^\mathcal{S}$ is a single convex polygon in plate space. For a cube-shaped planet, $\Omega_i$ is a clipped polyhedron and $\Omega_i^\mathcal{S}$ is the union of up to six convex polygons, one for each cube face the plate touches, each trimmed by the adjacent face planes.

In the implementation, each half-plane (or half-space) clip is performed in the manner of the Sutherland--Hodgman algorithm~\cite{sutherland1974}. Recall that this geometric representation is not identical to the pointwise representation due to the warp term $\mathbf{w}$ in Equation~\eqref{eq:warped-plate-space}. The two borders differ by at most $\sqrt{2}\,a_w/s_p$ world units, which Section~\ref{sec:river-node-segment-generation} uses to keep rivers inside their pointwise plate.

\subsection{Low-Resolution Plate Grid}

The river generator samples pre-river terrain using the weight fields on a low-resolution grid restricted to the geometric representation domain, providing approximate terrain elevation and slope for river path planning. Using a low-resolution terrain sample makes plate-local river preprocessing computationally practical by significantly reducing computation time and memory requirements, while still retaining a terrain-first generation philosophy.

For an infinite world, the implementation lays an $n_g\times n_g$ grid over the axis-aligned bounding rectangle of $\Omega_i^\mathcal{S}$ and keeps the samples that fall inside the polygon. For a cube-shaped planet, the grid covers each cube face that the plate touches and is cropped to that face's polygon. Each accepted grid sample $(u,v)$ on plate $i$ lies in the geometric surface domain and stores an attenuated planning height:
\begin{equation}
    \begin{gathered}
    \mathbf{x}_{uv} \in \Omega_i^\mathcal{S}, \qquad
    h_{uv}=H_{\mathrm{plan}}(\mathbf{x}_{uv}), \\
    \begin{aligned}
    H_{\mathrm{plan}} ={}& C + a_{\mathrm{land}}W_{\mathrm{land}} \\
    &+ a_{\mathrm{relief}}\left(W_{\mathrm{mountain}}+W_{\mathrm{plateau}}+W_{\mathrm{island}}\right),
    \end{aligned}
    \end{gathered}
\end{equation}
where $a_{\mathrm{land}}\le1$ and $a_{\mathrm{relief}}<1$ are pre-river attenuation factors ($a_{\mathrm{land}}=1$ and $a_{\mathrm{relief}}=0.35$ in the implementation), and the island noise of Equation~\eqref{eq:pre-river-height} is omitted. Because $H_{\mathrm{plan}}$ flattens mountains and plateaus relative to $H_{\mathrm{pre}}$, the river elevations derived from it generally lie below the surrounding terrain surface, which produces valley-like depressions after composition.

Each sample also records its distance to the domain border $\partial\Omega_i^\mathcal{S}$. A sample is \emph{border-safe} if this distance is at least the border margin $\omega_{\mathrm{border}}$, and \emph{lake-safe} if it is at least $\omega_{\mathrm{border}}+r_{\mathrm{lake}}+\omega_{\mathrm{river}}$, where $r_{\mathrm{lake}}$ is the lake radius (Section~\ref{sec:lazy-per-plate-river-cache}).

\subsection{River Node \& Segment Generation}
\label{sec:river-node-segment-generation}

\paragraph{Sources.} Candidate sources are border-safe grid samples with $h_{uv}\ge\max(h_{\mathrm{src}},h_{\mathrm{sea}})$. Candidates are visited in descending height order. A candidate is skipped if it lies within $s_{\min}$ grid cells of any river already accepted on the plate, including that river's source; otherwise it is traced immediately. Selection stops once $K$ paths have been accepted. Because this ordering depends only on the grid heights, source selection is deterministic. Keeping sources away from existing rivers means that every path begins at a new node, so paths share nodes only at confluences, and no new segment can overlap an existing one.

\paragraph{Tracing.} Each source is traced downhill one step at a time. From the current sample, the next sample is chosen within a $(2k+1)\times(2k+1)$ grid window: it must be valid, border-safe, and lower by at least a minimum drop $\varepsilon>0$, and the new segment must not cross any existing segment. Among these, the sample minimizing $h_{uv}+\beta\|\Delta_{uv}\|^2$ is selected, where the small weight $\beta$ favors shorter steps. The trace terminates when one of the following occurs:
\begin{itemize}
    \item the current sample is at or below sea level, producing a \emph{sea outlet};
    \item an existing non-source node within $k$ cells is lower by at least $\varepsilon$ and reachable without crossing, or the chosen step lands on an existing node, producing a \emph{confluence};
    \item no admissible lower sample exists, producing a \emph{lake} at a local minimum, provided that sample is lake-safe;
    \item the step budget $n_{\max}$ is exhausted.
\end{itemize}
If a dead end is not lake-safe, or if the only descent available would enter the border band, the entire path is discarded rather than truncated. Rivers therefore never drain into a plate border.

\paragraph{Monotone River Heights.} The raw heights $h$ strictly decrease along every accepted path by construction. Each path $(n_0,\ldots,n_m)$ is then assigned river heights $H$ by an affine remapping between its source and outlet. The outlet height $H_m$ is its raw height, or, if the path ends at a confluence, the already-assigned river height of the node it joins. The source height is $H_0=\max(h_0-\Delta_{\mathrm{drop}},\,H_m+\varepsilon)$, and
\begin{equation}
    H_k = H_m + \left(H_0-H_m\right)\frac{h_k-h_m}{h_0-h_m}, \qquad 0\le k\le m.
\end{equation}
Since $h_k$ is strictly decreasing in $k$, so is $H_k$. Paths are processed in generation order, so each tributary is fitted to the river height of the trunk it joins. Because paths share nodes only at confluences, no later refit changes an earlier path's heights, and heights agree at every confluence. Every stored segment is therefore strictly downhill. Finally, Strahler stream orders~\cite{strahler1957} are computed over the finished network and stored per segment: source tributaries receive order one, and the order increases where two equal-order tributaries meet.

Note that the algorithm for placing nodes and segments is arbitrary and can be modified or replaced without affecting the rest of the pipeline. However, to avoid blending discontinuities during final terrain composition, nodes and segments should not be placed near plate borders. The pointwise plate border differs from the geometric one by at most $\sqrt{2}\,a_w/s_p$ world units (Section~\ref{sec:geometric-representation}). The border margin is therefore chosen to exceed this displacement, $\omega_{\mathrm{border}} > \sqrt{2}\,a_w/s_p$. Every grid sample a river visits is border-safe, and the geometric domain is convex, so every point on every channel lies at least $\omega_{\mathrm{border}}$ inside it and is owned pointwise by the river's own plate. River-terrain blending extends a further $\omega_{\mathrm{river}}$ from a channel (see Section~\ref{sec:final-terrain}), and the lake-safe margin accounts for the lake radius and this width. Keeping the blend band itself inside the pointwise plate would require $\omega_{\mathrm{border}} \ge \sqrt{2}\,a_w/s_p + \omega_{\mathrm{river}}$; with a smaller margin, the outer edge of the band can be clipped near a border, which affects the carved valley but never the channel itself. The margin therefore trades against the warp amplitude: a stronger warp produces more irregular plate borders but requires a wider margin, which leaves less of each plate available for rivers. In the implementation, $a_w=0.01$ and $s_p=1.5\times10^{-4}$, giving a maximum displacement of about 94 world units against a margin of $\omega_{\mathrm{border}}=100$.

\subsection{Lazy Per-Plate River Cache}
\label{sec:lazy-per-plate-river-cache}

A plate's river network is generated the first time a terrain sample falls on that plate and finds the cache absent. When a view or chunk request is made, the caches of the $(2\rho_a+1)^2$ lattice plates around the plate owning the view center may also be built up front, for an active radius $\rho_a$, so that dense sampling does not stall on first contact. Caches are keyed by the plate's integer lattice index. Each cache stores the planning grid, the river nodes (with raw height, river height, and type), the segments (with endpoint river heights and Strahler order), and the accepted paths. Because rivers naturally should not cross tectonic boundaries---where plateaus and oceans generate within this system---each plate's network can be generated and cached in isolation without triggering generation of neighboring plates.

For a terrain sample on plate $i$, let $\mathcal{R}_i$ be the cached set of river segments. Lakes are treated as disks of radius $r_{\mathrm{lake}}$ centered on local-minimum nodes, so the per-segment distance is
\begin{equation}
    \delta_s(\mathbf{x}) =
    \min\!\Big(\operatorname{dist}(\mathbf{x},s),\;
    \min_{\mathbf{p}\in L_s}\|\mathbf{x}-\mathbf{p}\|-r_{\mathrm{lake}}\Big),
\end{equation}
where $L_s$ is the set of lake endpoints of $s$ (and the inner term is omitted when $L_s$ is empty). The distance to the river network is computed exactly as $\delta_{\mathrm{river}}(\mathbf{x})=\min_{s\in\mathcal{R}_i}\delta_s(\mathbf{x})$ and is non-positive inside a lake. No spatial index or cached nearest-segment regions are used: every query scans the segments of the owning plate, at a cost of $O(|\mathcal{R}_i|)$ per sample.

Each segment $s$ also provides a height $h_s(\mathbf{x})$, obtained by linearly interpolating its endpoint river heights at the clamped projection of $\mathbf{x}$ onto $s$. Within a margin $m_{\mathrm{lake}}$ of a lake, $h_s$ is set to the lake height, and beyond that margin it fades back toward the interpolated value over a distance $D_{\mathrm{lake}}$. The river height combines these per-segment heights:
\begin{equation}
    \label{eq:river-height-blend}
    H_{\mathrm{river}}(\mathbf{x}) =
    \begin{cases}
    h_{\mathrm{lake}}, & \text{in a lake,} \\
    \operatorname{mean}_{s:\,d_s=0} h_s(\mathbf{x}), & \text{on a channel,} \\
    \dfrac{\sum_{s} w_s h_s(\mathbf{x})}{\sum_{s} w_s}, & \text{otherwise,}
    \end{cases}
\end{equation}
with $d_s=\operatorname{dist}(\mathbf{x},s)$ and $w_s=\big(1-S(d_s/D_h)\big)^2/d_s^2$ over segments with $d_s<D_h$. If no segment lies within $D_h$, $h_s$ of the nearest segment is used. Compared with taking the height of the single nearest segment, this inverse-distance blend removes height discontinuities where the nearest segment changes, such as around bends and confluences. On the channel of any segment $s$ outside a lake surface, $H_{\mathrm{river}}=h_s$. Segments never cross or overlap, so the only other segments at zero distance then share an endpoint with $s$ and have the same height there. Because the lake fade only pulls $h_s$ toward the lower lake height as a channel approaches its lake, $H_{\mathrm{river}}$ never increases downstream along such a channel and is flat only on lake surfaces. Collinear overlaps cannot occur: a tracing step that touches the interior of an existing segment counts as a crossing and is rejected, and no source lies within $s_{\min}$ of an existing river.

The cached Strahler order can supply a river-size field, for example by taking the order of the nearest segment, and small deterministic noise offsets may be applied to the river distance, height, or size fields before final composition to make channels appear more curved. Neither affects the downhill guarantee, and neither is used in the height composition of the evaluated implementation.

\subsection{On-Demand Final Terrain Composition}
\label{sec:final-terrain}

During chunk generation, plate weight fields are sampled, the distance $\delta_{\mathrm{river}}(\mathbf{x})$ to the owning plate's river network is computed, and river elevation $H_{\mathrm{river}}(\mathbf{x})$ is evaluated by Equation~\eqref{eq:river-height-blend}. The river distance is converted into a normalized terrain blend weight before mixing the local terrain height with the river height:

\begin{equation}
    \begin{aligned}
    \lambda_{\mathrm{river}}(\mathbf{x}) &=
    \left[
    \min\!\left(1,\frac{[\delta_{\mathrm{river}}(\mathbf{x})]_+}{\omega_{\mathrm{river}}}\right)
    \right]^{\alpha_{\mathrm{river}}}, \\
    H_{\mathrm{final}}(\mathbf{x}) &=
    \lambda_{\mathrm{river}}(\mathbf{x})H_{\mathrm{pre}}(\mathbf{x})
    +
    \left(1-\lambda_{\mathrm{river}}(\mathbf{x})\right)H_{\mathrm{river}}(\mathbf{x}).
    \end{aligned}
\end{equation}
Here $\omega_{\mathrm{river}}$ is the river blending width and $\alpha_{\mathrm{river}}$ controls the blend falloff. Near the river, $\lambda_{\mathrm{river}}$ approaches zero and the interpolated river height dominates; away from the river, $\lambda_{\mathrm{river}}$ approaches one and the locally generated terrain is preserved. Because $\delta_{\mathrm{river}}\le0$ inside a lake, $\lambda_{\mathrm{river}}=0$ there, producing a flat lake surface. On every channel $\lambda_{\mathrm{river}}=0$ and $H_{\mathrm{final}}=H_{\mathrm{river}}$, and every channel point is owned pointwise by the channel's own plate (Section~\ref{sec:river-node-segment-generation}), so the downhill property of Section~\ref{sec:lazy-per-plate-river-cache} carries over to the final terrain along every channel.

\section{Experimental Results and Evaluation}

A video demonstration for a finite, cube-shaped planet implementation of this system is available on YouTube at \href{https://www.youtube.com/watch?v=4Ai_13znvgg}{youtube.com/watch?v=4Ai\_13znvgg}.

Code for the toy examples in Figure~\ref{fig:global-continent-rivers} is available at \href{https://github.com/Davismk3/plate-local-river-generation}{github.com/Davismk3/plate-local-river-generation}.

Figure~\ref{fig:global-continent-rivers}(a) and Table~\ref{tab:figure-export-metrics} show the output and measured performance of a single deterministic configuration of the flat infinite world, detailed below. Figures~\ref{fig:global-continent-rivers}(b) and~(c) illustrate the finite-world variants; no measurements are reported for them. All benchmark code and raw results accompany this manuscript on GitHub, and every statistic is computed directly from the released implementation.

The implementation is written in C++17 and has no third-party dependencies. It consists of a library for pointwise terrain evaluation, plate-local river construction, and lazy per-plate caching, together with a command-line heightmap exporter and a native interactive viewer. No game engine is used. The figure and metrics are produced by a separate exporter program that uses only the library's public interface and links zlib for PNG output. It renders the terrain with a software rasterizer at 3$\times$ supersampling and depth-tests river channels against the terrain. All code was compiled with Apple clang 17 at \texttt{-O3}.

\subsection{Reference Configuration}
\label{sec:reference-configuration}

A single exporter invocation produces the figure, a JSON summary, and a per-plate CSV file, so all reported measurements correspond to Figure~\ref{fig:global-continent-rivers}(a). The run uses seed 1 and a square view two plate-cell widths across, centered on plate $(1,0)$. The terrain is sampled at a resolution of $320\times320$ (102,400 samples). Ten plates own at least one sample in the view, and a river cache is built for each. River networks are generated on a $100\times100$ grid within each plate, with at most $K=50$ river paths per plate and a source spacing of $s_{\min}=5$ grid cells. The warp amplitude is $a_w=0.01$ plate cells, so the pointwise and geometric plate borders differ by at most about 94 world units. The border margin is $\omega_{\mathrm{border}}=100$ world units, the river blending width is also 100 world units, the lake radius is 150 world units, and the planning attenuation is $a_{\mathrm{relief}}=0.35$. Channel lengths are normalized to plate-cell widths.

The sampled terrain contains 63.92\% land and 36.08\% ocean at a sea level of 0.35. Terrain heights range from 0.200 to 1.435, with a median of 0.472 and a 95th percentile of 1.201.

The 10 plate caches contain 1,656 nodes and 1,450 segments. Of the 500 possible river paths, 222 are accepted (44.4\%); each plate accepts 18--28 paths before it runs out of source candidates that are both border-safe and at least $s_{\min}$ from an existing river. The networks contain 222 source nodes, 200 sea outlets, 6 lakes, 11 confluences, and 206 connected components, and the highest Strahler order is 2. The total channel length across all plate caches is 49.766 plate-cell widths, or 4.977 per plate on average. Individual plates contain 108--235 segments, and their total channel lengths range from 3.852 to 7.734 plate-cell widths.

After clipping the river network to the meshed domain shown in Figure~\ref{fig:global-continent-rivers}(a), 782 nodes and 655 segments remain. The visible network contains 130 source nodes, 81 sea outlets, 6 lakes, and 565 channel nodes, with a visible channel length of 22.582 plate-cell widths.

\subsection{Downhill Verification}
\label{sec:downhill-verification}

Downhill flow is verified at two levels. First, every stored river segment is tested using its adjusted endpoint river heights. All 1,450 segments descend, as guaranteed by construction (Section~\ref{sec:river-node-segment-generation}).

Second, the final composed terrain $H_{\mathrm{final}}$ is sampled at 16 evenly spaced points along every segment (23,200 samples), and any downstream rise greater than $10^{-9}$ is counted. No segment rises at any sample. Every one of these samples is owned pointwise by its segment's own plate, and no two segments overlap, confirming the guarantees of Sections~\ref{sec:river-node-segment-generation}--\ref{sec:final-terrain}.

Because the border margin equals the blending width rather than exceeding the warp displacement plus that width, the blend band around a channel is not guaranteed to stay within its plate. Sampling the band at $\pm\omega_{\mathrm{river}}/2$ and $\pm\omega_{\mathrm{river}}$ on either side of every channel sample finds 6 of 92,800 points (0.0065\%) owned by a neighboring plate. At those points, the outer edge of the carved valley is clipped; the channels themselves are unaffected.

\subsection{Memory and Timing}
\label{sec:memory-timing}

The cached nodes, segments, and paths of the 10 plates occupy 0.169~MiB (177,360 bytes), or approximately 122 bytes per river segment, at 64 bytes per node record and 40 bytes per segment record. Each cache also retains its planning grid, which occupies 4.10~MiB across the 10 plates. The grid is not needed for terrain queries and could be discarded after construction. Allocator and container overhead are excluded from both figures.

The measurements were collected on an Apple M4 Max system. Timings are the median of 5 repeats, each on a fresh world instance so that every cache is built cold; heights and networks were identical across repeats. Construction of the 10 plate caches requires 37.9~ms in total, with a mean of 3.81~ms and a median of 3.05~ms per cache, and individual caches take 2.33 to 10.07~ms. Sampling the $320\times320$ view with river queries requires 116~ms (1.13~$\mu$s per sample), compared with 28.1~ms (0.27~$\mu$s per sample) for pointwise terrain alone. The difference, about 0.85~$\mu$s per sample, is the cost of the exhaustive segment scan of Section~\ref{sec:lazy-per-plate-river-cache}. Rendering at 3$\times$ supersampling requires 0.161~s and PNG encoding 0.522~s. These timings are observational rather than controlled benchmarks and may vary with system load, whereas the terrain and river network statistics are deterministic.

\begingroup
\small
\refstepcounter{table}
\noindent\textbf{Table~\thetable.} Metrics emitted by the same deterministic exporter invocation that produced Figure~\ref{fig:global-continent-rivers}(a). Lengths are measured in plate-cell widths. Complete-network counts include all 10 plate caches intersecting the view; visible counts are clipped to the displayed mesh. Timings are medians of 5 cold repeats.\label{tab:figure-export-metrics}
\vspace{0.4em}

\tablefirsthead{\toprule \textbf{Metric} & \textbf{Value} \\ \midrule}
\tablehead{\multicolumn{2}{@{}l}{\textit{Table~\ref{tab:figure-export-metrics} continued}}\\ \toprule \textbf{Metric} & \textbf{Value} \\ \midrule}
\tabletail{\midrule \multicolumn{2}{@{}r}{\textit{continued on next column}}\\}
\tablelasttail{\bottomrule}
\begin{supertabular}{@{}lr@{}}
    Terrain samples & 102,400 \\
    Visible plate caches & 10 \\
    Land / ocean fraction & 63.92 / 36.08\% \\
    Median / 95th-percentile height & 0.472 / 1.201 \\
    Nodes / segments & 1,656 / 1,450 \\
    Accepted / possible paths & 222 / 500 \\
    Sources / confluences & 222 / 11 \\
    Sea outlets / lakes & 200 / 6 \\
    Connected components & 206 \\
    Maximum Strahler order & 2 \\
    Stored-segment downhill violations & 0 \\
    Final-terrain rising segments & 0 / 1,450 \\
    Channel samples on other plates & 0 / 23,200 \\
    Blend-band samples on other plates & 6 / 92,800 \\
    Total / visible channel length & 49.766 / 22.582 \\
    Visible nodes / segments & 782 / 655 \\
    Network cache size & 0.169 MiB \\
    Planning-grid size & 4.10 MiB \\
    Cache build total / mean & 37.9 / 3.81 ms \\
    Sample with / without rivers & 1.13 / 0.27 $\mu$s \\
    Render / PNG encode & 0.161 / 0.522 s \\
\end{supertabular}
\endgroup

\FloatBarrier

\section{Future Work}

River queries currently scan every segment of the owning plate. A per-plate spatial index, such as a uniform bin grid or bounding-volume hierarchy over the segments, would reduce this cost for plates with dense networks without changing the exact distance.

The plate-local framework is not limited to rivers. The same bounded-domain pattern could generate roads, towns, or other features that guarantee a certain topology. Each application would replace downhill flow with its own constraints---for example slope and travel cost for roads, or water access and terrain suitability for towns---while retaining deterministic identifiers, lazy regional preprocessing, per-plate caches, and pointwise evaluation.

Most notably, integration with learned generators such as InfiniteDiffusion~\cite{goslin2026} offers a longer-term hybrid direction. Plate-local fields and river graphs could supply explicit structural constraints, while a learned model adds fine surface, coastline, and mountain detail.

\section{Conclusion}

We have presented an algorithm for lazily generating river networks that are terrain-aware, deterministic, and coherent, with downhill flow guaranteed along every river channel of the final terrain, and which can be tailored to flat infinite world-types or closed-shaped finite world-types such as planets.

\end{document}